\documentclass[11pt]{article}
\usepackage[totalwidth=460truept,totalheight=600truept]{geometry}
\usepackage{latexsym,amssymb,amsmath,graphicx}
\usepackage[T1]{fontenc}

\def\theequation{\arabic{section}.\arabic{equation}}

\renewcommand{\theequation}{\thesection.\arabic{equation}}
\global\arraycolsep=1truept

\begin{document}

\hfill IFUP-TH 2008/18

\vskip 1.4truecm

\begin{center}
{\huge \textbf{Weighted Power Counting}}

{\huge \textbf{\large \vskip .1truecm}}

{\huge \textbf{And Lorentz Violating Gauge Theories.}}

{\huge \textbf{\large \vskip .1truecm}}

{\huge \textbf{II: Classification}}

\vskip 1.5truecm

\textsl{Damiano Anselmi}

\textit{Dipartimento di Fisica ``Enrico Fermi'', Universit\`{a} di Pisa, }

\textit{Largo Pontecorvo 3, I-56127 Pisa, Italy, }

\textit{and INFN, Sezione di Pisa, Pisa, Italy}

damiano.anselmi@df.unipi.it

\vskip 2truecm

\textbf{Abstract}
\end{center}

\bigskip

{\small We classify the local, polynomial, unitary gauge theories that
violate Lorentz symmetry explicitly at high energies and are renormalizable
by weighted power counting. We study the structure of such theories and
prove that renormalization does not generate higher time derivatives. We
work out the conditions to renormalize vertices that are usually
non-renormalizable, such as the two scalar-two fermion interactions and the
four fermion interactions. A number of four dimensional examples are
presented.}

\vskip 1truecm

\vfill\eject

\section{Introduction}

\setcounter{equation}{0}

Lorentz symmetry is a fundamental assumption behind the Standard Model of
particle physics. Experimental bounds on the parameters of the Lorentz
violating Standard-Model extension \cite{kostelecky} are often very precise 
\cite{koste}. Nevertheless, several authors, inspired by different
considerations, have suggested that Lorentz symmetry and CPT could be broken
at very high energies \cite{coletive}. The problem of Lorentz violation has
attracted a lot of interest, in cosmology, astrophysics, high-energy
physics. If Lorentz symmetry were not exact our understanding of Nature
would change considerably.

We can imagine that the Standard Model is corrected by Lorentz violating
terms of higher dimensions, multiplied by inverse powers of a scale $\Lambda
_{L}$, which can be understood as the scale of Lorentz violation. If $%
\Lambda _{L}$ is sufficiently large, the corrected model can be organized so
that it agrees with all present experimental data, yet it predicts
violations of Lorentz symmetry starting from energies $\sim \Lambda _{L}$.

If we do not assume exact Lorentz invariance at arbitrarily high energies,
yet demand locality and unitarity, several theories that are not
renormalizable by the usual power counting become renormalizable in the
framework of a ``weighted power counting'' \cite{renolor}, which assigns
different weights to space and time. The large momentum behavior of
propagators is improved by quadratic terms containing higher space
derivatives. The set of vertices is arranged so that no higher time
derivatives are generated by renormalization, in agreement with
(perturbative) unitarity. Scalar and fermion theories of this type have been
studied in \cite{renolor,confnolor}. In ref. \cite{LVgauge1suA}, to which we
refer as ``paper I'' from now on, the basic properties of Lorentz violating
gauge theories have been derived. Here we give an exhaustive classification
of gauge theories, investigate their structure and study a number of four
dimensional examples.

We search for theories that are local and polynomial, free of infrared
divergences in the Feynman diagrams, and renormalizable by weighted power
counting. To avoid the presence of certain spurious subdivergences,
originated by the peculiar form of the gauge-field propagator, spacetime is
split into space and time and other restrictions are imposed.

The paper is organized as follows. In section 2 we review the weighted power
counting. In section 3 we study the structure of renormalizable theories
containing gauge fields and matter. We work out the conditions for
polynomiality and renormalizability, and prove that higher time derivatives
are absent. In section 4 we derive sufficient conditions for the absence of
spurious subdivergences. In section 5 we study the conditions to renormalize
vertices that are usually non-renormalizable, in particular the two
scalar-two fermion interactions and the four fermion interactions, and
illustrate a number of four dimensional examples. In section 6 we consider
the most general type of Lorentz violations. Section 7 contains our
conclusions. In appendix A we recall the form of the gauge-field propagator
and the dispersion relations. In appendix B we study the renormalizability
of our theories to all orders, using the Batalin-Vilkovisky formalism.

\section{Weighted power counting}

\setcounter{equation}{0}

In this section we review the weighted power counting criterion of refs. 
\cite{renolor,confnolor} and a number of results from paper I. The simplest
framework to study the Lorentz violations is to assume that the $d$%
-dimensional Lorentz group $O(1,d-1)$ is broken to a residual symmetry $O(1,%
\hat{d}-1)\times O(\bar{d})$. The $d$-dimensional spacetime manifold $M=%
\mathbb{R}^{d}$ is split into the product $\hat{M}\times \bar{M}$ of two
submanifolds, a $\hat{d}$-dimensional submanifold $\hat{M}=\mathbb{R}^{\hat{d%
}}$, containing time and possibly some space coordinates, and a $\bar{d}$%
-dimensional space submanifold $\bar{M}=\mathbb{R}^{\bar{d}}$. The partial
derivative $\partial $ is decomposed as $(\hat{\partial},\bar{\partial})$,
where $\hat{\partial}$ and $\bar{\partial}$ act on the subspaces $\hat{M}$
and $\bar{M}$, respectively. Coordinates, momenta and spacetime indices are
decomposed similarly. We first study renormalization in this simplified
framework and later generalize our results to more general breakings (see
section 6). For simplicity, we assume separate invariances under C, P and T
throughout this paper.

Consider a free scalar theory with (Euclidean) lagrangian 
\begin{equation}
\mathcal{L}_{\hbox{free}}=\frac{1}{2}(\hat{\partial}\varphi )^{2}+\frac{1}{%
2\Lambda _{L}^{2n-2}}(\bar{\partial}^{n}\varphi )^{2},  \label{free}
\end{equation}
where $\Lambda _{L}$ is an energy scale and $n$ is an integer $>1$. It is
invariant under the weighted rescaling 
\begin{equation}
\hat{x}\rightarrow \hat{x}\ \mathrm{e}^{-\Omega },\qquad \bar{x}\rightarrow 
\bar{x}\ \mathrm{e}^{-\Omega /n},\qquad \varphi \rightarrow \varphi \ 
\mathrm{e}^{\Omega (\text{\dj }/2-1)},  \label{scale}
\end{equation}
where \dj $=\hat{d}+\bar{d}/n$ is the ``weighted dimension''. Note that $%
\Lambda _{L}$ is not rescaled.

The interacting theory is defined as a perturbative expansion around (\ref
{free}). For the purposes of renormalization, the masses and the other
quadratic terms can be treated perturbatively, since the counterterms depend
polynomially on them. Denote the ``weight'' of an object $\mathcal{O}$ by $[%
\mathcal{O}]$ and assign weights to coordinates, momenta and fields as
follows: 
\begin{equation}
\lbrack \hat{x}]=-1,\qquad [\bar{x}]=-\frac{1}{n},\qquad [\hat{\partial}%
]=1,\qquad [\bar{\partial}]=\frac{1}{n},\qquad [\varphi ]=\frac{\text{\dj }}{%
2}-1,  \label{weights}
\end{equation}
while $\Lambda _{L}$ is weightless. The lagrangian terms of weight \dj\ are
strictly renormalizable, those of weights smaller than \dj\
super-renormalizable and those of weights greater than \dj\
non-renormalizable. The weighted power counting criterion amounts to demand
that the theory contains no parameter of negative weight. The considerations
just recalled are easily generalized to fermions, whose weight is (\dj $%
-1)/2 $.

The gauge field $A_{\mu }=A_{\mu }^{a}T^{a}$, with $T^{a}$ anti-Hermitian,
is decomposed as $A=(\hat{A},\bar{A})$. The covariant derivative 
\begin{equation}
D=(\hat{D},\bar{D})=(\hat{\partial}+g\hat{A},\bar{\partial}+g\bar{A})
\label{3a}
\end{equation}
induces the weight assignments 
\[
\lbrack g\hat{A}]=[\hat{D}]=1,\qquad [g\bar{A}]=[\bar{D}]=\frac{1}{n}, 
\]
where $g$ is the gauge coupling. On the other hand, the weight-\dj\ kinetic
term $\sim (\hat{\partial}\hat{A})^{2}$ gives $[\hat{A}]=$\dj $/2-1$, so $%
[g]=2-$\dj $/2$. The field strength is split as 
\begin{equation}
\hat{F}_{\mu \nu }\equiv F_{\hat{\mu}\hat{\nu}},\qquad \tilde{F}_{\mu \nu
}\equiv F_{\hat{\mu}\bar{\nu}},\qquad \bar{F}_{\mu \nu }\equiv F_{\bar{\mu}%
\bar{\nu}}.  \label{3b}
\end{equation}
We find 
\begin{equation}
\lbrack \hat{A}]=\frac{\text{\dj }}{2}-1,\qquad [\bar{A}]=\frac{\text{\dj }}{%
2}-2+\frac{1}{n},\qquad [\hat{F}]=\frac{\text{\dj }}{2},\qquad [\tilde{F}]=%
\frac{\text{\dj }}{2}-1+\frac{1}{n},\qquad [\bar{F}]=\frac{\text{\dj }}{2}-2+%
\frac{2}{n}.  \label{we}
\end{equation}

In the presence of gauge interactions the renormalizable theories are still
those that do not contain parameters of negative weights. To single out the
super-renormalizable theories we can refine this requirement, demanding that
no parameters have weights smaller than some non-negative constant $\chi $.
Indeed, if that happens Feynman diagrams are certainly multiplied by
coefficients of weights greater than or equal to $\chi $, so no new
counterterms are turned on by renormalization. Applying the refined
requirement to the gauge coupling $g$ we find 
\begin{equation}
0\leq \chi \leq 2-\text{\dj }/2.  \label{chi}
\end{equation}
In particular, we must have \dj $\leq 4$.

It is convenient to write the gauge-field action 
\begin{equation}
\mathcal{S}_{0}=\int \mathrm{d}^{d}x\left( \mathcal{L}_{Q}+\mathcal{L}%
_{I}\right) \equiv \mathcal{S}_{Q}+\mathcal{S}_{I},  \label{s0}
\end{equation}
as the sum of two contributions $\mathcal{S}_{Q}$ and $\mathcal{S}_{I}$: $%
\mathcal{S}_{Q}$ collects the gauge-invariant quadratic terms of weight $%
\leq $\dj , constructed with two field strengths and possibly covariant
derivatives, while $\mathcal{S}_{I}$ collects the vertex terms of weights $%
\leq $\dj $-\chi $, constructed with at least three field strengths and
possibly covariant derivatives.

Up to total derivatives the quadratic part $\mathcal{L}_{Q}$ of the
lagrangian reads (in the Euclidean framework) 
\begin{equation}
\mathcal{L}_{Q}=\frac{1}{4}\left\{ \hat{F}_{\mu \nu }^{2}+2F_{\hat{\mu}\bar{%
\nu}}\eta (\bar{\Upsilon})F_{\hat{\mu}\bar{\nu}}+F_{\bar{\mu}\bar{\nu}}\tau (%
\bar{\Upsilon})F_{\bar{\mu}\bar{\nu}}+\frac{1}{\Lambda _{L}^{2}}(D_{\hat{\rho%
}}F_{\bar{\mu}\bar{\nu}})\xi (\bar{\Upsilon})(D_{\hat{\rho}}F_{\bar{\mu}\bar{%
\nu}})\right\} .  \label{l0}
\end{equation}
The proof can be found in paper I. Here $\bar{\Upsilon}\equiv -\bar{D}%
^{2}/\Lambda _{L}^{2}$ and $\eta $, $\tau $ and $\xi $ are polynomials of
degrees $n-1$, $2n-2$ and $n-2$, respectively. We have expansions 
\begin{equation}
\eta (\bar{\Upsilon})=\sum_{i=0}^{n-1}\eta _{n-1-i}\bar{\Upsilon}^{i},\qquad
[\eta _{j}]=\frac{2j}{n},  \label{expo}
\end{equation}
and similar, where $\eta _{i}$ are dimensionless constants of non-negative
weights.

The free action is positive definite if and only if 
\begin{equation}
\eta >0,\qquad \tilde{\eta}\equiv \eta +\frac{\bar{k}^{2}}{\Lambda _{L}^{2}}%
\xi >0,\qquad \tau >0,  \label{pos}
\end{equation}
where now $\eta $, $\tau $ and $\xi $ are functions of $\bar{k}^{2}/\Lambda
_{L}^{2}$. Furthermore, we assume 
\begin{equation}
\eta _{0}>0,\qquad \tau _{0}>0,\qquad \tilde{\eta}_{0}=\eta _{0}+\xi
_{0}>0,\qquad \eta _{n-1}>0,\qquad \tau _{2n-2}>0.  \label{bilu}
\end{equation}
The first three conditions ensure that the propagators have the best UV
behaviors. The other two conditions, together with 
\begin{equation}
d\geq 4,  \label{irabse}
\end{equation}
ensure that the Feynman diagrams are free of IR\ divergences at
non-exceptional external momenta, despite the fact that the gauge fields are
massless. The reason is that, under the mentioned assumptions, the IR\
behavior of Feynman diagrams is governed by the low-energy theory 
\begin{equation}
\mathcal{L}_{\text{IR}}=\frac{1}{4}\left[ (F_{\hat{\mu}\hat{\nu}%
}^{a})^{2}+2\eta _{n-1}(F_{\hat{\mu}\bar{\nu}}^{a})^{2}+\tau _{2n-2}(F_{\bar{%
\mu}\bar{\nu}}^{a})^{2}\right] ,  \label{ir}
\end{equation}
which has an ordinary power counting.

The BRST\ symmetry \cite{brs} coincides with the usual one, 
\begin{eqnarray*}
sA_{\mu }^{a} &=&D_{\mu }^{ab}C^{b}=\partial _{\mu }C^{a}+gf^{abc}A_{\mu
}^{b}C^{c},\qquad sC^{a}=-\frac{g}{2}f^{abc}C^{b}C^{c}, \\
s\bar{C}^{a} &=&B^{a},\qquad sB^{a}=0,\qquad s\psi
^{i}=-gT_{ij}^{a}C^{a}\psi ^{j},
\end{eqnarray*}
etc., with the weight assignments 
\begin{equation}
\lbrack C]=[\bar{C}]=\frac{\text{\dj }}{2}-1,\qquad [s]=1,\qquad [B]=\frac{%
\text{\dj }}{2}.  \label{g1}
\end{equation}

We choose the gauge-fixing 
\begin{equation}
\mathcal{L}_{\text{gf}}=s\Psi ,\qquad \Psi =\bar{C}^{a}\left( -\frac{\lambda 
}{2}B^{a}+\mathcal{G}^{a}\right) ,\qquad \mathcal{G}^{a}\equiv \hat{\partial}%
\cdot \hat{A}^{a}+\zeta \left( \bar{\upsilon}\right) \bar{\partial}\cdot 
\bar{A}^{a},  \label{gf}
\end{equation}
where $\lambda $ is a dimensionless, weightless constant, $\bar{\upsilon}%
\equiv -\bar{\partial}^{2}/\Lambda _{L}^{2}$ and $\zeta $ is a polynomial of
degree $n-1$. Compatibly with (\ref{bilu}) we assume 
\begin{equation}
\zeta >0,\qquad \zeta _{0}>0,\qquad \zeta _{n-1}>0.  \label{gfpos}
\end{equation}

The total gauge-fixed action is 
\begin{equation}
\mathcal{S}=\int \mathrm{d}^{d}x\left( \mathcal{L}_{Q}+\mathcal{L}_{I}+%
\mathcal{L}_{\text{gf}}\right) \equiv \mathcal{S}_{0}+\mathcal{S}_{\text{gf}%
}.  \label{basis}
\end{equation}
The propagator is reported in appendix A, together with the dispersion
relations.

For the purposes of renormalization, we can treat the weightful parameters $%
\eta _{i}$, $\tau _{i}$, $\xi _{i}$ and $\zeta _{i}$, $i>0$, perturbatively,
because the divergent parts of Feynman diagrams depend polynomially on them.
In this framework, the propagators we use in the high-energy analysis of the
diagrams are (\ref{pros})-(\ref{fg2}) with the replacements 
\[
\eta \rightarrow \eta _{0}\left( \frac{\bar{k}^{2}}{\Lambda _{L}^{2}}\right)
^{n-1},\qquad \tau \rightarrow \tau _{0}\left( \frac{\bar{k}^{2}}{\Lambda
_{L}^{2}}\right) ^{2(n-1)},\qquad \xi \rightarrow \xi _{0}\left( \frac{\bar{k%
}^{2}}{\Lambda _{L}^{2}}\right) ^{n-2},\qquad \zeta \rightarrow \zeta
_{0}\left( \frac{\bar{k}^{2}}{\Lambda _{L}^{2}}\right) ^{n-1}, 
\]
every other term being treated as a vertex. Intermediate masses can be added
to the denominators, to avoid IR\ problems, and removed immediately after
calculating the divergent parts.

We recall that $P_{k,n}(\hat{p},\bar{p})$ is a weighted polynomial in $\hat{p%
}$ and $\bar{p}$, of degree $k$, where $k$ is a multiple of $1/n$, if $%
P_{k,n}(\xi ^{n}\hat{p},\xi \bar{p})$ is a polynomial of degree $kn$ in $\xi 
$. A propagator is regular if it is the ratio 
\begin{equation}
\frac{P_{r}(\hat{k},\bar{k})}{P_{2s}^{\prime }(\hat{k},\bar{k})}
\label{uvbeha}
\end{equation}
of two weighted polynomials of degrees $r$ and $2s$, where $r$ and $s$ are
integers, such that the denominator $P_{2s}^{\prime }(\hat{k},\overline{k})$
is non-negative (in the Euclidean framework), non-vanishing when either $%
\hat{k}\neq 0$ or $\overline{k}\neq 0$ and has the form 
\begin{equation}
P_{s}^{\prime }(\hat{k},\bar{k})=\hat{\omega}(\hat{k}^{2})^{s}+\bar{\omega}(%
\bar{k}^{2})^{ns}+\cdots ,  \label{uvbeha2}
\end{equation}
with $\hat{\omega}>0$, $\bar{\omega}>0$, where the dots collect the terms $(%
\hat{k}^{2})^{j-m}(\bar{k}^{2})^{mn}$ with $j<s$, $0\leq m\leq j$, and $j=s$%
, $0<m<s$. The regularity conditions ensure that the derivatives with
respect to $\hat{k}$ improve the large-$\bar{k}$ behavior (because $\bar{%
\omega}\neq 0$), besides the large-$\hat{k}$ and overall ones, and the
derivatives with respect to $\bar{k}$ improve the large-$\hat{k}$ behavior
(because $\hat{\omega}\neq 0$), besides the large-$\bar{k}$ and overall
ones. For this reason, the $\hat{k}$-subdivergences are local in $\bar{k}$
and the $\bar{k}$-subdivergences are local in $\hat{k}$. The $\hat{k}$%
-subintegrals and the $\overline{k}$-subintegrals, which cannot behave worse
than the $\hat{k}$-$\overline{k}$-integrals, are automatically cured by the
counterterms that subtract the overall divergences of the $\hat{k}$-$%
\overline{k}$-integrals. Such counterterms are, for example, the first terms
of the ``weighted Taylor expansion'' around vanishing external momenta \cite
{renolor}.

A propagator that does not satisfy (\ref{uvbeha}) can generate spurious
ultraviolet subdivergences in Feynman diagrams when $\hat{k}$ tends to
infinity at $\overline{k}$ fixed, or viceversa. The gauge and ghost
propagators (\ref{pros}), (\ref{prosg}) are regular at non-exceptional
momenta, because the positivity conditions (\ref{pos}) and (\ref{gfpos})
ensure that the denominators are positive-definite in the Euclidean
framework. Moreover, the conditions (\ref{bilu}) ensure that all such
propagators but $\langle \bar{A}\bar{A}\rangle $ satisfy (\ref{uvbeha})-(\ref
{uvbeha2}) in the Feynman gauge (\ref{fg2}). Instead, $\langle \bar{A}\bar{A}%
\rangle $ is regular when $\overline{k}$ tends to infinity at $\hat{k}$
fixed, but not when $\hat{k}$ tends to infinity at $\overline{k}$ fixed,
where it behaves like $\sim 1/\hat{k}^{2}$. To ensure that no spurious
subdivergence is generated by the $\hat{k}$-subintegrals, a more careful
analysis must be performed, to which we devote section 4. The result is that
the sufficient conditions to ensure the absence of spurious subdivergences
include 
\begin{equation}
\hat{d}=1,\qquad d=\text{even}\qquad n=\text{odd},  \label{subd}
\end{equation}
plus other restrictions stated at the end of section 4. In particular,
spacetime is split into space and time. In section 6 we prove that, because
of the spurious subdivergences, more general type of Lorentz violations ($%
\hat{d}>1$) are disfavored.

The absence of spurious subdivergences ensures the locality of counterterms.
Consider a diagram $G_{r}$ equipped with the subtractions that take care of
its diverging proper subdiagrams. Differentiating $G_{r}$ a sufficient
number of times with respect to any components $\hat{p}_{i}$, $\bar{p}_{i}$
of the external momenta $p_{i}$, we can arbitrarily reduce the overall
degree of divergence and eventually produce a convergent integral.
Therefore, overall divergences are polynomial in all components of the
external momenta.

\section{Structure of renormalizable theories}

\setcounter{equation}{0}

In this section we investigate renormalizable and super-renormalizable
theories in detail. We study the conditions for renormalizability and
polynomiality, and investigate the time-derivative structure. In section 4
we study the spurious subdivergences, while section 5 is devoted to explicit
examples, mainly four dimensional.

We know that the theories contain only parameters of weights $\geq \chi $,
where $\chi $ satisfies (\ref{chi}). Call $\lambda _{i}$ the coupling
multiplying the $i$-th vertex belonging to the physical sector and denote
the number of its external legs by $n_{i}$. Clearly, $n_{i}\geq 3$ and $%
[\lambda _{i}]\geq \chi $. By polynomiality, the number of physical vertices
is finite, so we can take $\chi \equiv \min_{i}[\lambda _{i}]$. Define 
\[
\kappa \equiv \min_{i}\frac{[\lambda _{i}]}{n_{i}-2}. 
\]
Since the gauge coupling multiplies three-leg vertices, we have 
\begin{equation}
\lbrack \lambda _{i}]\geq (n_{i}-2)\kappa \quad \forall i,\qquad \text{and }%
0\leq \kappa \leq 2-\frac{\text{\dj }}{2},  \label{kappa}
\end{equation}
and $\chi >0$ if and only if $\kappa >0$. Introduce a coupling $\bar{g}$ of
weight $\kappa $ and write $\lambda _{i}=\bar{\lambda}_{i}\bar{g}^{n_{i}-2}$%
. Then (\ref{kappa}) ensures $[\bar{\lambda}_{i}]\geq 0$. The theory can be
reformulated in the ``$1/\bar{\alpha}$ form'' ($\bar{\alpha}=\bar{g}^{2}$),
namely as 
\begin{equation}
\mathcal{L}_{1/\bar{\alpha}}=\frac{1}{\bar{\alpha}}\mathcal{\bar{L}}_{r}(%
\bar{g}A,\bar{g}\varphi ,\bar{g}\psi ,\bar{g}\bar{C},\bar{g}C,\bar{\lambda}),
\label{sur2}
\end{equation}
where $\varphi $ and $\psi $ are matter fields (scalars and fermions,
respectively) and the reduced lagrangian $\mathcal{\bar{L}}_{r}$ depends
polynomially on $\bar{g}$ and the $\bar{\lambda}$'s. The gauge coupling can
be parametrized as $g=\bar{g}\rho $, where $\rho $ has a non-negative weight
and is included in the set of the $\bar{\lambda}$'s. A generic vertex of (%
\ref{sur2}) has the structure 
\begin{equation}
\bar{\lambda}_{i}\bar{g}^{n_{i}-2}\hat{\partial}^{k}\bar{\partial}^{m}\hat{A}%
^{p}\bar{A}^{q}\bar{C}^{r}C^{r}\varphi ^{s}\bar{\psi}^{t}\psi ^{t},
\label{vert}
\end{equation}
where $n_{i}=p+q+2r+s+2t$ and $p,q,r,k,m,s$ and $t$ are integers. Formula (%
\ref{vert}) and analogous expressions in this paper are meant
``symbolically'', which means that we pay attention to the field- and
derivative-contents of the vertices, but not where the derivatives act and
how Lorentz, gauge and other indices are contracted.

Every counterterm generated by (\ref{sur2}) fits into the structure (\ref
{sur2}). Indeed, consider a $L$-loop diagram with $E$ external legs, $I$
internal legs and $v_{i}$ vertices of type $i$. The leg-counting gives $%
\sum_{i}n_{i}v_{i}=E+2I=E+2(L+V-1)$, so the diagram is multiplied by a
product of couplings 
\begin{equation}
\bar{g}^{\sum_{i}(n_{i}-2)v_{i}}\prod_{i}\bar{\lambda}_{i}^{v_{i}}=\bar{%
\alpha}^{L}\bar{g}^{E-2}\prod_{i}\bar{\lambda}_{i}^{v_{i}}.  \label{ax}
\end{equation}
We see that a $\bar{g}^{E-2}$ factorizes, as expected. Moreover, each loop
order carries an additional weight of at least $2\kappa $.

When $\kappa =2-$\dj $/2$ we can take $\bar{g}=g$, which gives the $1/\alpha 
$ theories considered in paper I. They have a lagrangian of the form 
\begin{equation}
\mathcal{L}_{1/\alpha }=\frac{1}{\alpha }\mathcal{L}_{r}(gA,g\varphi ,g\psi
,g\bar{C},gC,\lambda ).  \label{sur}
\end{equation}
The class (\ref{sur2}) is much richer than the class (\ref{sur}), yet it is
does not cover the most general case.

To move a step forward towards the most general class of theories, it is
useful to show how to gauge scalar-fermion theories. Express the matter
theory in $1/\bar{\alpha}$ form, namely 
\begin{equation}
\mathcal{L}_{\text{matter}}=\frac{1}{\bar{\alpha}}\mathcal{\bar{L}}_{sf}(%
\bar{g}\varphi ,\bar{g}\psi ,\bar{\lambda}_{sf}).  \label{sf}
\end{equation}
We assume that $[g]\geq [\bar{g}]$ and write $g=\bar{g}\rho $, with $[\rho
]\geq 0$. In this way, the gauge interactions can be switched off letting $%
\rho $ tend to zero. Covariantize the derivatives contained in (\ref{sf})
and add the $1/\alpha $ pure gauge theory, plus extra terms allowed by the
weighted power counting. We obtain a mixed theory of the form 
\begin{equation}
\mathcal{L}=\frac{1}{\alpha }\mathcal{L}_{g}(gA,g\bar{C},gC,\lambda _{g})+%
\frac{1}{\bar{\alpha}}\mathcal{\bar{L}}_{sf}(\bar{g}\varphi ,\bar{g}\psi ,%
\bar{\lambda}_{sf})+\frac{1}{\bar{\alpha}}\Delta \mathcal{L}(gA,g\bar{C},gC,%
\bar{g}\varphi ,\bar{g}\psi ,\lambda ).  \label{str}
\end{equation}
Here $\Delta \mathcal{L}$ contains both the terms necessary to covariantize $%
\mathcal{\bar{L}}_{sf}$ and the mentioned extra terms. Consider a diagram $G$
with $E$ external legs and $L$ loops. Using the $\bar{g}$-$\rho $
parametrization and repeating the argument that leads to (\ref{ax}) we find
that $G$ is multiplied by $\bar{\alpha}^{L}\bar{g}^{E-2}$, so it agrees with
the structure (\ref{str}). On the other hand, every vertex of $\mathcal{L}%
_{g}$ used to construct $G$ provides at least two internal legs. Therefore,
every external $A$-, $\bar{C}$- and $C$-leg of $G$ is multiplied by at least
one power of $\rho $. This proves that the structure (\ref{str}) is
renormalizable. The theory is polynomial if $[\bar{g}\varphi ],[\bar{g}\psi
]>0$, namely \dj $>2-2\kappa $ if scalar fields are present, \dj $>1-2\kappa 
$ if the matter sector contains only fermions.

Now we are ready to introduce the most general class of theories, where
different fields can carry different $\bar{g}$'s. Call $\bar{g}_{i}$, $%
i=1,2,3,$ the ones of vectors, fermions and scalars, respectively\footnote{%
A more general situation where different subsets of fields with the same
spin have different $\bar{g}$'s is also possible. This generalization is
straightforward and left to the reader.}. As in (\ref{sur2}), $\bar{g}_{1}$
needs not coincide with $g$. Call $\bar{\gamma}_{k}$, $k=1,2,3$, the
coupling of minimum weight between $\bar{g}_{i}$ and $\bar{g}_{j}$, where $%
k\neq i,j$. Call $\bar{g}$ the coupling of minimum weight among the $\bar{g}%
_{i}$'s. Define $\bar{\alpha}_{i}=\bar{g}_{i}^{2}$, $\bar{a}_{i}=\bar{\gamma}%
_{i}^{2}$.

The lagrangian has the weight structure 
\begin{eqnarray}
\mathcal{L} &=&\frac{1}{\bar{\alpha}_{1}}\mathcal{L}_{1}(\bar{g}_{1}A)+\frac{%
1}{\bar{\alpha}_{2}}\mathcal{L}_{2}(\bar{g}_{2}\psi )+\frac{1}{\bar{\alpha}%
_{3}}\mathcal{L}_{3}(\bar{g}_{3}\varphi )+\frac{1}{\bar{a}_{3}}\mathcal{L}%
_{12}(\bar{g}_{1}A,\bar{g}_{2}\psi )  \nonumber \\
&&+\frac{1}{\bar{a}_{2}}\mathcal{L}_{13}(\bar{g}_{1}A,\bar{g}_{3}\varphi )+%
\frac{1}{\bar{a}_{1}}\mathcal{L}_{23}(\bar{g}_{2}\psi ,\bar{g}_{3}\varphi )+%
\frac{1}{\bar{\alpha}}\mathcal{L}_{123}(\bar{g}_{1}A,\bar{g}_{2}\psi ,\bar{g}%
_{3}\varphi ).  \label{mixed}
\end{eqnarray}
In $A$ we collectively include also ghosts and antighosts. Any other
parameters $\lambda $ contained in (\ref{mixed}) must have non-negative
weights. The $\bar{g}_{i}$-factors appearing in formula (\ref{mixed}) are
mere tools to keep track of the weight structure. For example, instead of $%
\bar{g}_{2}\psi $ we can have any $\bar{g}_{i}\psi $, as long as $[\bar{g}%
_{i}]\geq [\bar{g}_{2}]$. Similarly, the denominators $1/\bar{\alpha}_{i}$, $%
1/\bar{a}_{i}$ and $1/\bar{\alpha}$ are devices that lower the weights of
appropriate amounts.

Every $\mathcal{L}$ on the right-hand side of (\ref{mixed}) must be
polynomial in the fields and parameters. Moreover, we assume 
\begin{equation}
\lbrack g]\geq [\bar{g}_{1}],\qquad [g\bar{g}_{1}]\geq [\bar{g}%
_{2}^{2}],\qquad [g\bar{g}_{1}]\geq [\bar{g}_{3}^{2}].  \label{tu}
\end{equation}
These inequalities ensure that (\ref{mixed}) is compatible with the
covariant structure. Indeed, because of (\ref{tu}), the vertices generated
by covariant derivatives are multiplied by factors of weights not smaller
than the ones appearing in (\ref{mixed}), so they can fit into one of the
structures (\ref{mixed}). Observe that (\ref{tu}) implies $[g]\geq [\bar{g}%
_{i}]$ for every $i$.

Again, it is easy to prove that the structure (\ref{mixed}) is preserved by
renormalization. Assume, for example, that $[\bar{g}_{1}]\geq $ $[\bar{g}%
_{2}]\geq [\bar{g}_{3}]$ (the other cases can be treated symmetrically,
because (\ref{tu}) plays no role here) and write $\bar{g}_{1}=\rho \sigma 
\bar{g}$, $\bar{g}_{2}=\sigma \bar{g}$, $\bar{g}_{3}=\bar{g}$, with $[\rho
]\geq 0$, $[\sigma ]\geq 0$. In the parametrization $\bar{g}$-$\rho $-$%
\sigma $ the $\bar{g}$-powers in front of counterterms can be counted as in (%
\ref{ax}). Moreover, vertices contain a factor $\sigma $ for every $A$- and $%
\psi $-leg, save two legs in $\varphi $-independent vertices. Since at least
two legs of every vertex enter the diagrams, counterterms contain at least a
factor $\sigma $ for every external $A$- and $\psi $-leg. A similar argument
applies to $\rho $-factors and external $A$-legs. Thus, every diagram with $%
L\geq 1$ loops, $E$ external legs, $E_{A}$ external $A$-legs and $E_{\psi }$
external $\psi $-legs is multiplied at least by a factor $\bar{\alpha}^{L}%
\bar{g}^{E-2}\rho ^{E_{A}}\sigma ^{E_{A}+E_{\psi }}$ and therefore fits into
the structure (\ref{mixed}).

This argument proves also that the one-loop counterterms generated by (\ref
{mixed}) have the weight structure 
\begin{equation}
\Delta _{1}\mathcal{L}(\bar{g}_{1}A,\bar{g}_{2}\psi ,\bar{g}_{3}\varphi ),
\label{mixec}
\end{equation}
while at $L$ loops there is an additional factor of $\bar{\alpha}^{L-1}$.
Simplified versions of our theories can be obtained dropping vertices and
quadratic terms of (\ref{mixed}) that are not contained in (\ref{mixec}),
because renormalization is unable to generate them back. The quadratic terms
that cannot be dropped are those that control the behavior of propagators.
Of course, the simplified model must also contain the vertices related to
such quadratic terms by covariantization.

\paragraph{Polynomiality}

Now we derive the conditions to have polynomiality. Consider first the
physical (i.e. non gauge-fixing) sectors of the lagrangian (\ref{mixed}).
Apart from the factors $1/\bar{\alpha}_{i}$, $1/\bar{a}_{i}$ and $1/\bar{%
\alpha}$, they depend only on the products $\bar{g}_{1}F$, $\bar{g}_{2}\psi $%
, $\bar{g}_{3}\varphi $, and their covariant derivatives, so polynomiality
is ensured when these objects have positive weights. Let us focus for the
moment on the gauge sector. From (\ref{we}) we see that if $\bar{d}>1$ the
most meaningful condition is $[\bar{g}_{1}\bar{F}]>0$. If instead $\bar{d}=1$
the most meaningful condition is $[\bar{g}_{1}\tilde{F}]>0$, because $\bar{F}%
\equiv 0$. However, because of (\ref{subd}) and (\ref{irabse}) we have to
concentrate on the former case. We conclude that pure gauge theories are
polynomial in the physical sector if and only if 
\[
4-\frac{4}{n}-2\kappa _{1}<\text{\dj }, 
\]
having written $[\bar{g}_{i}]=\kappa _{i}$. In the presence of scalars and
fermions we must have 
\begin{equation}
4-\frac{4}{n}-2\kappa _{1}<\text{\dj },\qquad 1-2\kappa _{2}<\text{\dj }%
,\qquad 2-2\kappa _{3}<\text{\dj .}  \label{h}
\end{equation}

Observe that (\ref{h}) and $n\geq 2$ ensure that the weight of $\bar{g}_{1}%
\hat{A}$ is strictly positive. Thus the theory is certainly polynomial in $%
\hat{A}$. For the same reason, it is polynomial also in $\bar{C}$ and $C$.
On the other hand, the weight of $\bar{g}_{1}\bar{A}$ can be negative,
because (\ref{h}) ensures only $[\bar{g}_{1}\bar{A}]>-1/n$. This means that,
in principle, the gauge-fixing sector can be non-polynomial. Now we show
that if the tree-level gauge fixing is (\ref{gf}), then the theory is
polynomial also in the gauge-fixing sector. Note that is some cases (see
appendix B) the gauge-fixing sector does not preserve the simple form (\ref
{gf}), but can acquire new types of vertices by renormalization.

We need to prove that beyond the tree level, in both the physical and
gauge-fixing sectors, the field $\bar{A}$ appears only in the combinations 
\begin{equation}
g\bar{A},\qquad \bar{g}_{1}\hat{\partial}\bar{A},\qquad \bar{g}_{1}g\hat{A}%
\bar{A},\qquad \bar{g}_{1}\bar{\partial}\bar{A},\qquad \bar{g}_{1}g\bar{A}%
\bar{A}.  \label{tipa}
\end{equation}
First observe that at the tree level this statement is true up to the
factors $1/\bar{\alpha}_{i}$, $1/\bar{a}_{i}$ and $1/\bar{\alpha}$ appearing
in (\ref{mixed}). Indeed, $\bar{A}$ appears only in the following locations: 
$i$) in $(\mathcal{G}^{a})^{2}$, which contributes only to the propagator; $%
ii$) inside the covariant derivative $\bar{D}$ (also in the ghost action); $%
iii$) inside the field strength. In case $ii$) $\bar{A}$ is multiplied by $g$
and gives the first term of (\ref{tipa}). In case $iii$) the field strength
carries an extra factor $\bar{g}_{1}$: $\bar{g}_{1}\tilde{F}$ gives the
second and third terms of (\ref{tipa}), while $\bar{g}_{1}\bar{F}$ gives the
forth and fifth terms.

Next, consider an $L$-loop Feynman diagram $G$ and assume that the $\bar{A}$%
-structure of the renormalized action is (\ref{tipa}) up to the order $L-1$
included, with the tree-level caveat just mentioned. The factors $1/\bar{%
\alpha}_{i}$, $1/\bar{a}_{i}$ and $1/\bar{\alpha}$ of (\ref{mixed}) are
simplified by the internal legs of $G$, which are at least two for every
vertex. Consider the $\bar{A}$-external legs of $G$. In the first case of (%
\ref{tipa}) the $\bar{A}$-leg is accompanied by a factor $g$ and in the
second case by a $\bar{g}_{1}$ and a derivative $\hat{\partial}$ acting on
it. In the third case it carries a factor $g$ (the $\bar{g}_{1}$ being left
for the $\hat{A}$-leg, in case it is external), in the forth case a $\bar{g}%
_{1}$ and a derivative $\bar{\partial}$. In the fifth case both $\bar{A}$'s
or just one $\bar{A}$ can be external, with factors $\bar{g}_{1}g$ and $g$,
respectively. Therefore, diagrams and counterterms contain $\bar{A}$ only in
the combinations (\ref{tipa}), so the property (\ref{tipa}) is inductively
promoted to all orders.

Under the conditions (\ref{h}) $\bar{g}_{1}\hat{A}$ and each combination (%
\ref{tipa}) have positive weights. The $\bar{A}$ external legs are always
equipped with enough $\bar{g}_{1}$-$g$-factors and/or derivatives to raise
the weight by a finite amount. Thus, the total renormalized lagrangian is
polynomial, gauge-fixing sector included.

In conclusion, recalling (\ref{irabse}), (\ref{subd}) and (\ref{h}),
consistent renormalizable gauge theories with a non-trivial
super-renormalizable subsector require 
\begin{equation}
n=\text{odd},\qquad d=\text{even}\geq 4,\qquad \hat{d}=1,\qquad 4-\frac{4}{n}%
<\text{\dj }+2\kappa _{1},\qquad 1<\text{\dj }+2\kappa _{2},\qquad 2<\text{%
\dj }+2\kappa _{3}\text{,}  \label{d=42}
\end{equation}
plus other restrictions summarized at the end of section 4 to ensure the
absence of spurious subdivergences. Moreover, (\ref{tu}) gives 
\begin{equation}
\kappa _{1}\leq 2-\frac{\text{\dj }}{2},\qquad \kappa _{2,3}\leq 1+\frac{%
\kappa _{1}}{2}-\frac{\text{\dj }}{4},  \label{h2}
\end{equation}
and of course we must have $\kappa _{i}\geq 0$.

The same argument that leads to (\ref{tipa}) proves that the counterterms
contain the field $\hat{A}$ only in the combinations 
\begin{equation}
g\hat{A},\qquad \bar{g}_{1}\hat{\partial}\hat{A},\qquad \bar{g}_{1}g\hat{A}%
\hat{A},\qquad \bar{g}_{1}\bar{\partial}\hat{A},\qquad \bar{g}_{1}g\hat{A}%
\bar{A}.  \label{tipa2}
\end{equation}
Again, at the tree level this statement is true up to the factors $1/\bar{%
\alpha}_{i}$, $1/\bar{a}_{i}$ and $1/\bar{\alpha}$ appearing in (\ref{mixed}%
).

\paragraph{Time-derivative structure}

To ensure (perturbative) unitarity it is crucial to prove that the
lagrangian contains no terms with higher time derivatives. We now prove that
it is so and give a complete classification of the $\hat{\partial}$%
-structure.

Using the information encoded in (\ref{tipa}) and (\ref{tipa2}) a generic
lagrangian term can be schematically written as 
\begin{eqnarray}
&&\frac{\bar{\lambda}_{i}}{\bar{g}^{\prime 2}}(\hat{\partial}+g\hat{A})^{k}(%
\bar{\partial}+g\bar{A})^{m}(\bar{g}_{1}\hat{\partial}\hat{A}+\bar{g}_{1}g%
\hat{A}\hat{A})^{p}(\bar{g}_{1}\hat{\partial}\bar{A}+\bar{g}_{1}\bar{\partial%
}\hat{A}+\bar{g}_{1}g\hat{A}\bar{A})^{q}  \nonumber \\
&&(\bar{g}_{1}\bar{\partial}\bar{A}+\bar{g}_{1}g\bar{A}\bar{A})^{h}(\bar{g}%
_{1}^{2}\bar{C}C)^{r}(\bar{g}_{3}\varphi )^{s}(\bar{g}_{2}^{2}\bar{\psi}\psi
)^{t},  \label{nueva}
\end{eqnarray}
where $[\bar{\lambda}_{i}]\geq 0$ and $\bar{g}^{\prime }$ is the $\bar{g}$
of minimum weight among those appearing in the vertex. We find the
inequality 
\begin{eqnarray}
&k+\frac{m}{n}+(2p+q)\left( 1-\frac{1}{n}\right) +(p+q+h)\left( \frac{\text{%
\dj }}{2}-2+\frac{2}{n}+\kappa _{1}\right) &  \nonumber \\
&+r\left( \text{\dj }-2+2\kappa _{1}\right) +s\left( \frac{\text{\dj }}{2}%
-1+\kappa _{3}\right) +t\left( \text{\dj }-1+2\kappa _{2}\right) -\text{\dj }%
-2\kappa ^{\prime }\leq &0.  \label{yh}
\end{eqnarray}
Moreover, we know that $\kappa ^{\prime }=[\bar{g}^{\prime }]$ is not larger
than any of the other $\kappa $'s appearing in the inequality. Observe that
every quantity between parenthesis is non-negative.

First we study the vertices, then the quadratic terms. Consider the vertices
containing fermions ($t\geq 1$). We have two possibilities: $i$) $%
p=q=h=r=s=0 $, $t=1$ and $k+m/n\leq 1$; or $ii$) 
\[
k+\frac{m}{n}+(2p+q)\left( 1-\frac{1}{n}\right) <1. 
\]
Case $i$) gives no vertex with time derivatives. In case $ii$) we have
immediately $k=p=0$, $q\leq 1$. Time derivatives are contained only in terms
of the form 
\begin{equation}
X_{1}^{\prime }\equiv f_{1}^{\prime }(\bar{A},\varphi ,\psi ,\bar{C},C,\bar{%
\partial})(\hat{\partial}\bar{A}),  \label{u}
\end{equation}
where $\bar{\partial}$ can act anywhere.

From now on we can neglect the fermions. Consider the vertices with two or
more scalars. Again, we have two cases: $iii$) $p=q=h=r=0$, $s=2$ and $%
k+m/n\leq 2$; or $iv$) 
\begin{equation}
k+\frac{m}{n}+(2p+q)\left( 1-\frac{1}{n}\right) <2.  \label{due}
\end{equation}
In case $iii$)\ vertices can have at most one time derivative and fall in
the class 
\begin{equation}
X_{1}\equiv \hat{\partial}f_{1}(\hat{A},\bar{A},\varphi ,\bar{C},C,\bar{%
\partial}),  \label{x1}
\end{equation}
where the $\hat{\partial}$-and $\bar{\partial}$-derivatives are allowed to
act anywhere. In case $iv$) we must have $k\leq 1$. For $k=1$, we have
either $p=0$, $q=1$, $m=0$, which is not $O(\bar{d})$-invariant, or $p=q=0$,
which is not $O(1,\hat{d}-1)$-invariant. For $k=0$ we have $p=1$, $q=0$,
which is of the form (\ref{x1}), or $p=0$, $q=2$, which is of the form (\ref
{x1}) or 
\begin{equation}
X_{2}\equiv f_{2}(\bar{A},\varphi ,\bar{C},C,\bar{\partial})(\hat{\partial}%
\bar{A})(\hat{\partial}\bar{A}),  \label{vk1}
\end{equation}
where only the $\bar{\partial}$-derivatives can act anywhere.

Next, consider the vertices with one scalar. If $r\geq 1$ we have again (\ref
{due}), therefore vertices of the form (\ref{x1}) or (\ref{vk1}). If $r=0$
consider first the case $p+q+h\geq 2$. Then we have 
\begin{equation}
k+\frac{m}{n}+(2p+q-4)\left( 1-\frac{1}{n}\right) <0,  \label{mus}
\end{equation}
so either $p=1$ or $p=0$. If $p=1$ we can have only $k=q=0$, which has the
form (\ref{x1}). If $p=0$ we can have $k=1,0$. If $k=1$ then $q=1$, so the
vertex is of the form (\ref{x1}), (\ref{vk1}) or 
\begin{equation}
X_{2}^{\prime }\equiv f_{2}^{\prime }(\bar{A},\varphi ,\bar{\partial})(\hat{%
\partial}^{2}\bar{A}).  \label{vk2}
\end{equation}
If $k=0$ then $q\leq 2$ so the vertex is of the form (\ref{x1}) or (\ref{vk1}%
). It it easy to see that also the vertices with $s=1$, $r=0$ and $p+q+h<2$
fall in the classes (\ref{x1}), (\ref{vk1}) or (\ref{vk2}).

Now consider the vertices with neither scalars nor fermions. Here $\kappa
^{\prime }=\kappa _{1}$. If $r\geq 1$ we have either $v$) $p=q=h=0$, $r=1$;
or $vi$) (\ref{due}). These cases are the same as $iii$) and $iv$) above,
with two scalar fields replaced by ghosts. The vertices they give fall in
the classes listed so far. We remain with the vertices with $s=r=t=0$. We
have the cases: $vii$) $p+q+h>2$; $viii$) $p+q+h\leq 2$. In case $vii$) we
have (\ref{mus}) again, therefore $k\leq 3$, $2p+q<4$. If $k=3$ then $q=1$,
by $O(1,\hat{d}-1)$ invariance, but it violates (\ref{mus}). If $k=2$ we
have $p=q=0$, while if $k=1$ we have $q=1$, $p=0$. In either case the
vertices fall in the classes (\ref{x1}), (\ref{vk2}) and (\ref{vk2}). If $%
k=0 $ then $p=1$, $q=0$, which gives (\ref{x1}), or $p=0$, $q=2$, which
gives (\ref{x1}) or (\ref{vk1}). It is easy to show that case $viii$) does
not produce new types of vertices with time derivatives.

Finally, the quadratic terms that do not fall in the classes (\ref{x1}), (%
\ref{vk1}) and (\ref{vk2}) have the forms 
\begin{equation}
(\hat{\partial}\hat{A})^{2},\qquad \bar{C}\hat{\partial}^{2}C,\qquad \varphi 
\hat{\partial}^{2}\varphi ,\qquad \bar{\psi}\hat{\partial}\psi ,
\label{quad}
\end{equation}
as expected. Every other term is $\hat{\partial}$-independent. We conclude,
in particular, that the theory is free of higher time derivatives.

\section{Absence of spurious subdivergences}

\setcounter{equation}{0}

In this section we derive sufficient conditions to ensure the absence of
spurious subdivergences. We generalize the proof given in paper I, which was
specific for $1/\alpha $ theories. We use the Feynman gauge (\ref{fg2}) and
the dimensional-regularization technique. We proceed by induction and assume
that counterterms corresponding to diverging proper subdiagrams are
included. Moreover, we assume $\hat{d}=1$, $n=$odd and that the spacetime
dimension is even. We also assume that the theory does not contain vertices
of type $X_{1}^{\prime }$ , see \ref{u}), which is true in most physical
applications. Other restrictions will emerge along with the analysis. The
complete sets of sufficient conditions are recapitulated at the end of the
section.

Consider a generic $N$-loop integral

\begin{equation}
\int \frac{\mathrm{d}\hat{k}_{1}}{(2\pi )^{\hat{d}}}\int \frac{\mathrm{d}^{%
\bar{d}}\bar{k}_{1}}{(2\pi )^{\bar{d}}}\cdots \int \frac{\mathrm{d}\hat{k}%
_{N}}{(2\pi )^{\hat{d}}}\int \frac{\mathrm{d}^{\bar{d}}\bar{k}_{N}}{(2\pi )^{%
\bar{d}}},  \label{a1}
\end{equation}
with loop momenta $(k_{1},\ldots ,k_{N})$. We have to prove that all
subintegrals, in all parametrizations $(k_{1}^{\prime },\ldots
,k_{N}^{\prime })$ of the momenta, are free of subdivergences. By the
inductive assumption, all divergent subintegrals 
\begin{equation}
\prod_{j=1}^{M}\int \frac{\mathrm{d}\hat{k}_{j}^{\prime }}{(2\pi )^{\hat{d}}}%
\int \frac{\mathrm{d}^{\bar{d}}\bar{k}_{j}^{\prime }}{(2\pi )^{\bar{d}}},
\label{a2}
\end{equation}
where $M<N$, are subtracted by appropriate counterterms. We need to consider
subintegrals where some hatted integrations are missing and the
corresponding barred integrations are present, and/or viceversa.

The proof given in paper I is divided in three steps: structure of
integrals, $\hat{k}$--subintegrals and mixed subintegrals. The first step
does not have to be repeated here, since it applies unchanged. It proves
that we can focus on the subintegrals containing some $\hat{k}_{a}^{\prime }$%
-integrations without the corresponding $\bar{k}_{a}^{\prime }$%
-integrations. We generalize the second step of the proof and the third one.

\paragraph{$\hat{k}$-subintegrals}

Now we prove that the subintegrals over hatted components of momenta have no
spurious subdivergences. More precisely, we prove, under very general
assumptions, that one-dimensional integrals have no logarithmic divergences,
namely their renormalization-group flow is trivial. This property ensures
that using the dimensional-regularization technique, which kills the
power-like divergences automatically, the $\hat{k}$-subintegrals are
convergent.

Consider ``Feynman integrals'' in one dimension, and assume that:\ $i$) the
propagators are regular everywhere; $ii$) when $p$ is large they behave as $%
1/(p^{2})^{N}$ times some polynomial in $p$, for some $N<\infty $; $iii$)
they tend to a constant for $p\rightarrow 0$. In Lorentz violating gauge
theories such assumptions hold with $N$ equal to 1, but our proof is more
general. Consider a diagram $G$ with $L$ loops, $V$ vertices and $I$
internal legs. Denote the loop momenta with $p_{i}$. We have an integral
that for large $p_{i}$'s looks like 
\[
\mathcal{I}(L,V,\omega )=\int \prod_{i=1}^{L}\mathrm{d}p_{i}\frac{P_{\omega
}^{\prime }(p)}{\left( \prod_{j=1}^{L}(p_{i}^{2})^{N}\right) P_{V-1}\left(
(\Delta p^{2})^{N}\right) }. 
\]
We have used $I=L+V-1$. Here $P_{V-1}$ is a polynomial of degree $V-1$ in $%
(\Delta p^{2})^{N}$, where $\Delta p$ are linear combinations of the $p$'s
with coefficients $\pm 1$. The numerator $P_{\omega }^{\prime }$ is a
polynomial of degree $\omega $ in the $p$'s. To have a potential overall
divergence we need 
\begin{equation}
\omega \geq L(2N-1)+2N(V-1).  \label{pu}
\end{equation}
If $V=1$ the integral factorizes into $L$ one-loop integrals, which cannot
contain logarithmic divergences. Assume $V>1$. Then (\ref{pu}) implies that
each monomial of $P_{\omega }^{\prime }(p)$ contains at least $2N$ powers of
some $p_{i}$, say $p_{1}$, which ``simplify'' a propagator. Actually they
produce a regular function of the form 
\begin{equation}
\frac{p_{1}^{2N}}{p_{1}^{2N}+\sum_{j=1}^{2N}c_{j}p_{1}^{2N-j}}=1-\frac{%
\sum_{j=1}^{2N}c_{j}p_{1}^{2N-j}}{p_{1}^{2N}+\sum_{j=1}^{2N}c_{j}p_{1}^{2N-j}%
}.  \label{tre}
\end{equation}
Consider first the ``1'' on the right-hand side of this equation. It gives 
\[
\int \prod_{i=1}^{L}\mathrm{d}p_{i}\frac{P_{\omega -2N}^{\prime }(p)}{\left(
\prod_{j=2}^{L}(p_{i}^{2})^{N}\right) P_{V-1}\left( (\Delta
p^{2})^{N}\right) }. 
\]
We can distinguish two cases: $i$) $P_{V-1}$ does not depend on $p_{1}$; $ii$%
) $P_{V-1}$ depends on $p_{1}$. In case $i$) the $p_{1}$-integral factorizes
and cannot produce logarithmic divergences. We remain with a $\mathcal{I}%
(L-1,V,\omega -2N)$. In case $ii$), after a $p_{1}$-translation we obtain an
integral $\mathcal{I}(L,V-1,\omega -2N)$. The translation can cost at worst
another $\mathcal{I}(L-1,V,\omega -2N)$. Now consider the second term on the
right-hand side of (\ref{tre}): it gives $iii$) a $\mathcal{I}(L,V,\omega
^{\prime })$ with $\omega ^{\prime }<\omega $.

In all cases we can repeat the arguments made so far, with fewer loops or
vertices, or with a smaller $\omega $. At each step either an integral
factorizes, or a propagator simplifies, or $\omega $ decreases. We end up
with zero loops, namely no integral, or one vertex, namely $L$ factorized
integrals, or an $\omega $ violating (\ref{pu}), i.e. an overall convergent
integral. Proceeding this way we find that there cannot be logarithmic
divergences. If there are no logarithmic divergences at $\hat{d}=1$ there
are no divergences at all reaching $\hat{d}=1$ from complex dimensions $%
1-\varepsilon _{1}$.

\paragraph{Mixed subintegrals}

Consider subintegrals of the form 
\begin{equation}
\prod_{i=1}^{L}\int \frac{\mathrm{d}\hat{k}_{i}^{\prime }}{(2\pi )^{\hat{d}}}%
\left[ \prod_{j=L+1}^{L+M}\int \frac{\mathrm{d}\hat{k}_{j}^{\prime }}{(2\pi
)^{\hat{d}}}\int \frac{\mathrm{d}^{\bar{d}}\bar{k}_{j}^{\prime }}{(2\pi )^{%
\bar{d}}}\right] ,  \label{a4}
\end{equation}
which are ``incomplete'' in $L$ barred directions. The complete subintegrals
in square brakets can be regarded as products of (nonlocal, but one-particle
irreducible) ``subvertices''. Let $r$ label the subvertices with $\tilde{n}_{%
\hat{A}r}$, $\tilde{n}_{\bar{A}r}$, $\tilde{n}_{Cr}$, $\tilde{n}_{fr}$, $%
\tilde{n}_{sr}$ external legs of types $\hat{A}$, $\bar{A}$, ghost, fermion
and scalar, respectively. Since subvertices are at least one-loop, each leg
has a factor $\bar{g}_{1,2,3}$ attached to it (see (\ref{mixec})). Thus, the
weight $\tilde{\delta}_{r}$ of the subvertices of type $r$ satisfies the
bound 
\begin{eqnarray}
\tilde{\delta}_{r} &\leq &\text{\dj }-(\tilde{n}_{\hat{A}r}+\tilde{n}%
_{Cr})\left( \frac{\text{\dj }}{2}-1+\kappa _{1}\right) -\tilde{n}_{\bar{A}%
r}\left( \frac{\text{\dj }}{2}-2+\frac{1}{n}+\kappa _{1}\right)  \nonumber \\
&&-\tilde{n}_{fr}\left( \frac{\text{\dj }-1}{2}+\kappa _{2}\right) -\tilde{n}%
_{sr}\left( \frac{\text{\dj }}{2}-1+\kappa _{3}\right) .  \label{deltatilde}
\end{eqnarray}

Consider a subintegral (\ref{a4}) corresponding to a subdiagram with $L$
loops, $v_{1}$ vertices of type $X_{1}$, $v_{2}$ vertices of type $X_{2}$
and $X_{2}^{\prime }$, $\Delta v_{i}$ vertices of other types with $2i$
fermionic legs, $I_{B}$ internal bosonic legs (including ghosts) and $I_{F}$
internal fermionic legs, $E_{F}$ external fermionic legs and $\tilde{v}_{r}$
subvertices of type $r$. We know that every bosonic propagator behaves at
least like $1/\hat{k}^{2}$, for $\hat{k}$ large, while the fermionic
propagator behaves like $1/\hat{k}$. The subintegral has a superficial
degree of divergence equal to 
\begin{equation}
\hat{\omega}(G)=L+v_{1}+2v_{2}-2I_{B}-I_{F}+\sum_{r}\tilde{v}_{r}\tilde{%
\delta}_{r}.  \label{b1}
\end{equation}
Moreover, the topological identity $L-I+V=1$ gives 
\begin{equation}
L=1+I_{B}+I_{F}-v_{1}-v_{2}-\sum_{i=0}^{i_{\text{max}}}\Delta v_{i}-\sum_{r}%
\tilde{v}_{r}.  \label{b2}
\end{equation}
Counting the fermionic legs of the subdiagram we have 
\begin{equation}
2I_{F}+E_{F}=\sum_{i=0}^{i_{\text{max}}}2i\Delta v_{i}+\sum_{r}\tilde{v}_{r}%
\tilde{n}_{fr}.  \label{b3}
\end{equation}
Combining (\ref{b1}), (\ref{b2}) and (\ref{b3}) we get 
\begin{equation}
\hat{\omega}(G)=2-L-v_{1}-\frac{E_{F}}{2}+\sum_{i=0}^{i_{\text{max}%
}}(i-2)\Delta v_{i}+\sum_{r}\tilde{v}_{r}\left( \tilde{\delta}_{r}-2+\frac{%
\tilde{n}_{fr}}{2}\right) .  \label{degre}
\end{equation}
We know that in the realm of the usual power counting, odd-dimensional
integrals do not have logarithmic divergences. In the realm of the weighted
power counting, such a property generalizes as follows: if $\hat{d}=1$, $d=$%
even and $n=$odd, then odd-dimensional (weighted) integrals do not have
logarithmic divergences. The proof is simple and left to the reader. Thus,
the case $L=1$ is excluded. Sufficient conditions to have $\hat{\omega}%
(G)\leq 0$ are then $i_{\text{max}}\leq 2$ (i.e. vertices with more than
four fermionic legs should be absent) and 
\begin{equation}
\tilde{\delta}_{r}-2+\frac{1}{2}\tilde{n}_{fr}<0\qquad \text{for every }r%
\text{.}  \label{soll}
\end{equation}
Indeed, if such conditions hold (\ref{degre}) gives $\hat{\omega}(G)<0$
unless all subvertices are absent, which is the case considered previously.
Using (\ref{deltatilde}), sufficient conditions to have (\ref{soll}) are 
\begin{equation}
\text{\dj }-(\tilde{n}_{\hat{A}r}+\tilde{n}_{Cr})c_{1}-\tilde{n}_{\bar{A}r}%
\bar{c}_{1}-\tilde{n}_{fr}c_{2}-\tilde{n}_{sr}c_{3}<2,  \label{soll2}
\end{equation}
where 
\begin{equation}
c_{i}=\frac{\text{\dj }}{2}-1+\kappa _{i},\qquad \bar{c}_{1}=\frac{\text{\dj 
}}{2}-2+\frac{1}{n}+\kappa _{1}.  \label{cis}
\end{equation}

Finally, the most general mixed subintegrals have the form 
\[
\prod_{i=1}^{L}\int \frac{\mathrm{d}\hat{k}_{i}^{\prime }}{(2\pi )^{\hat{d}}}%
\left[ \prod_{j=L+1}^{L+M}\int \frac{\mathrm{d}\hat{k}_{j}^{\prime }}{(2\pi
)^{\hat{d}}}\int \frac{\mathrm{d}^{\bar{d}}\bar{k}_{j}^{\prime }}{(2\pi )^{%
\bar{d}}}\prod_{m=L+M+1}^{L+M+P}\int \frac{\mathrm{d}^{\bar{d}}\bar{k}%
_{m}^{\prime }}{(2\pi )^{\bar{d}}}\right] . 
\]
They can be treated as above, considering the integrals between square
brakets as subvertices. Now formula (\ref{degre}) has an extra $-P$ on the
right-hand side, since $P$ hatted intergations are missing. The situation,
therefore, can only improve. The only caveat is that now $L$ can also be one
(if $P$ is odd). Even in that case, however, $2-L-P\leq 0$, since $P\geq 1$.

\paragraph{Sufficient conditions for the absence of spurious subdivergences}

Now we work out sufficient conditions to fulfill (\ref{soll}). First, we
require that the coefficients $c_{i}$, $\bar{c}_{1}$ in (\ref{soll2}) be
strictly positive, so that the bound (\ref{soll2}) improves when the number
of legs increases. It is easy to check that under such conditions vertices $%
X_{1}^{\prime }$ and vertices with more than four fermionic legs are
automatically forbidden. We can have various cases, according to which of
the $c_{i}$, $\bar{c}_{1}$ is minimum. $i$) If 
\begin{equation}
2-\frac{1}{n}-\frac{\text{\dj }}{2}<\kappa _{1}\leq \kappa _{2,3}+1-\frac{1}{%
n}  \label{b4}
\end{equation}
the minimum coefficient is $\bar{c}_{1}$. Then the worst case for the bound (%
\ref{deltatilde}) is $\tilde{n}_{\bar{A}r}=2$, $\tilde{n}_{\hat{A}r}=\tilde{n%
}_{Cr}=\tilde{n}_{fr}=\tilde{n}_{sr}=0$, so sufficient conditions to ensure (%
\ref{soll2}) are 
\begin{equation}
1-\frac{1}{n}<\kappa _{1}.  \label{b5}
\end{equation}
The combination of (\ref{b4}) and (\ref{b5}) is 
\begin{equation}
1-\frac{1}{n}<\kappa _{1}\leq \kappa _{2,3}+1-\frac{1}{n},\qquad \kappa
_{1}>2-\frac{1}{n}-\frac{\text{\dj }}{2}.  \label{b6}
\end{equation}
Repeating the argument for the other cases, we find $ii$)\ 
\begin{equation}
1-\frac{\text{\dj }}{2}<\kappa _{2}\leq \kappa _{3},\qquad 1-\frac{1}{n}%
+\kappa _{2}\leq \kappa _{1},  \label{b6p}
\end{equation}
with minimum coefficients $c_{2}$, and $iii$) (\ref{b6p}) with $\kappa _{2}$
and $\kappa _{3}$ interchanged, with minimum coefficient $c_{3}$.

The case \dj $\leq 2$ is important for physical applications, so we treat it
apart. We prove that sufficient conditions for the absence of spurious
subdivergences are 
\begin{equation}
\text{\dj }\leq 2,\qquad \kappa _{1}>2-\frac{1}{n}-\frac{\text{\dj }}{2}%
,\qquad \kappa _{2}\geq 1-\frac{\text{\dj }}{2}.  \label{b7}
\end{equation}
The second and third inequalities of this list ensure $\bar{c}_{1}>0$ and $%
c_{2}\geq 0$, respectively, while $c_{3}>0$ is already ensured by the last
inequality of (\ref{d=42}). Again, it is easy to prove that under such
conditions neither vertices $X_{1}^{\prime }$, nor vertices with more than
four fermionic legs are allowed. Moreover, four fermion vertices cannot have
other types of legs.

Now, (\ref{b7}) imply $\hat{\omega}(G)\leq 0$. If \dj $<2$ or $\kappa
_{2}>1- $\dj $/2$, then $\hat{\omega}(G)=0$ only for $\tilde{v}_{r}=0$.
Instead, if \dj $=2$ and $\kappa _{2}=0$, then $\hat{\omega}(G)=0$ for $%
\tilde{v}_{r}=0$ or 
\[
L=2-P,\quad v_{1}=E_{F}=\Delta v_{0}=\Delta v_{1}=\tilde{n}_{\hat{A}r}=%
\tilde{n}_{\bar{A}r}=\tilde{n}_{Cr}=\tilde{n}_{sr}=0,\quad \tilde{n}%
_{fr},v_{2},\Delta v_{2}=\text{arbitrary.} 
\]
The subdiagrams with such features do not contain vertices with both
fermionic and bosonic legs, have no external fermionic leg and their
subvertices have only fermionic legs. Thus, either $\tilde{v}_{r}=0$ and the
diagram falls in the $\hat{k}$-subintegral class discussed above, or $%
v_{2}=0 $ and the diagram has no external leg, therefore it is trivial.

Concluding, if $\hat{d}=1$, $n=$odd and $d$ is even spurious subdivergences
are absent if either (\ref{b6}), or (\ref{b6p}), or (\ref{b6p}) with $\kappa
_{2}\leftrightarrow \kappa _{3}$, or (\ref{b7}) hold.

\section{Renormalizable theories}

\setcounter{equation}{0}

In this section we study examples of renormalizable theories and look for
four dimensional models that contain two scalar-two fermion interactions and
four fermion interactions.

The simplest models are those that have the smallest values of $n$ ($\geq 2$%
) and the largest values of $\kappa _{1,2,3}$. Compatibly with (\ref{h2})
the largest value of all $\kappa _{i}$'s is $2-$\dj $/2$, which gives the $%
1/\alpha $ theories considered in paper I. Those models exhibit, in a
simplified framework, several properties of Lorentz violating gauge
theories, but are not particularly rich from a phenomenological point of
view, because they cannot renormalize vertices that are otherwise
non-renormalizable, apart from those that contain some unusual dependences
on $\bar{A}$ and $\bar{\partial}$.

The simplest four dimensional $1/\alpha $ theory \cite{LVgauge1suA} has $n=2$%
, \dj $=5/2$ and the (Euclidean) lagrangian 
\begin{equation}
\mathcal{L}_{1/\alpha }=\mathcal{L}_{Q}+\frac{g}{\Lambda _{L}^{2}}%
f_{abc}\left( \lambda \tilde{F}_{\hat{\mu}\bar{\nu}}^{a}\tilde{F}_{\hat{\mu}%
\bar{\rho}}^{b}+\lambda ^{\prime }\bar{F}_{\bar{\mu}\bar{\nu}}^{a}\bar{F}_{%
\bar{\mu}\bar{\rho}}^{b}\right) \bar{F}_{\bar{\nu}\bar{\rho}}^{c}+\frac{g}{%
\Lambda _{L}^{4}}\sum_{j}\lambda _{j}\bar{D}^{2}\bar{F}^{3}{}_{j}+\frac{%
\alpha }{\Lambda _{L}^{4}}\sum_{k}\lambda _{k}^{\prime }\bar{F}^{4}{}_{k},
\label{de}
\end{equation}
where $j$ labels the independent gauge invariant terms constructed with two
covariant derivatives $\bar{D}$ acting on three field strengths $\bar{F}$,
and $k$ labels the terms constructed with four $\bar{F}$'s. The last two
terms are symbolic.

Let us investigate the $1/\bar{\alpha}$ extensions of (\ref{de}). The
maximal extension is the one with $\kappa =0$. The theory contains the
additional vertices 
\begin{equation}
\sum_{p=2}^{4}\frac{\lambda _{p}}{\Lambda _{L}^{9p/2}}\tilde{F}^{2}\bar{F}%
^{p}+\sum_{q=4}^{6}\frac{\lambda _{q}^{\prime }}{\Lambda _{L}^{9q/2-7}}\bar{D%
}^{2}\bar{F}^{q}+\sum_{r=5}^{10}\frac{\lambda _{r}^{\prime \prime }}{\Lambda
_{L}^{9r/2-9}}\bar{F}^{r}.  \label{yu}
\end{equation}
Larger values of $\kappa $ can reduce the set of vertices in various ways.
For example, for $5/12<\kappa \leq 3/4$ the theory is still (\ref{de}). For $%
1/4<\kappa \leq 5/12$ we have a unique additional vertex, $\bar{F}^{5}$. For 
$3/20<\kappa \leq 1/4$ we have also $\tilde{F}^{2}\bar{F}^{2}$, $\bar{D}^{2}%
\bar{F}^{4}$ and $\bar{F}^{6}$. For $1/12<\kappa \leq 3/20$ we have also $%
\bar{F}^{7}$, and so on. However, because $n$ is even the model (5.1) and
its extensions (5.2) may have spurious subdivergences. Going through the
previous section it is possible to show that such subdivergences appear only
at three loops. The first completely consistent model is thus the theory
with $\hat{d}=1$, $n=3$, \dj $=2$. Its simplest renormalizable lagrangian is
the sum of $\mathcal{L}_{Q}$ plus $\bar{F}^{3}$.

\paragraph{Theories with two scalar-two fermion vertices}

Two scalar-two fermion interactions

\[
\frac{\bar{g}_{2}^{2}\bar{g}_{3}^{2}}{\bar{a}_{1}}\varphi ^{2}\bar{\psi}\psi 
\]
are renormalizable if and only if 
\[
\kappa _{2,3}\leq \frac{3}{2}-\frac{\text{\dj }}{2} 
\]
and of course $\kappa _{2,3}\geq 0$. Let us choose the largest values of $%
\kappa _{1,2,3}$ compatible with this bound and (\ref{d=42})-(\ref{h2}),
namely 
\[
\bar{g}_{1}=g,\qquad \bar{g}_{2}=\bar{g}_{3}=\bar{g},\qquad \kappa
_{2}=\kappa _{3}=\frac{3}{2}-\frac{\text{\dj }}{2}. 
\]

We can take, for example, $n=2$, \dj $=5/2$, with gauge group $SU(2)\times
U(1)$ and matter fields in the fundamental representation of $SU(2)$. Then
we have the theory 
\begin{eqnarray}
\mathcal{L} &=&\mathcal{L}_{1/\alpha }+\bar{\psi}\left( \hat{D}\!\!\!\!\slash%
+\frac{\eta _{f}}{\Lambda _{L}}{\bar{D}\!\!\!\!\slash}^{\,2}+\eta
_{f}^{\prime }{\bar{D}\!\!\!\!\slash}+m_{f}\right) \psi +  \nonumber \\
&&+|\hat{D}\varphi |^{2}+\frac{\eta _{s}}{\Lambda _{L}^{2}}|\bar{D}%
^{2}\varphi |^{2}+\eta _{s}^{\prime }|\bar{D}\varphi |^{2}+m_{s}^{2}|\varphi
|^{2}+\frac{\lambda _{4}\bar{g}^{2}}{4}|\varphi |^{4}  \nonumber \\
&&+\frac{\bar{g}^{2}}{4\Lambda _{L}^{2}}\left[ \frac{\lambda _{6}}{9}\bar{g}%
^{2}|\varphi |^{6}+\lambda _{4}^{(3)}|\varphi |^{2}|\bar{D}\varphi
|^{2}+\lambda _{4}^{(2)}|\varphi ^{\dagger }\bar{D}\varphi |^{2}+\lambda
_{4}^{(1)}\left( (\varphi ^{\dagger }\bar{D}\varphi )^{2}+\text{h.c.}\right)
\right]  \nonumber \\
&&+\frac{Y\bar{g}^{2}}{\Lambda _{L}}|\varphi |^{2}\bar{\psi}\psi +\frac{%
Y^{\prime }\bar{g}^{2}}{\Lambda _{L}}(\bar{\psi}\varphi )(\varphi ^{\dagger
}\psi )+\frac{\tau _{f}g}{\Lambda _{L}}i{\bar{F}}_{\bar{\mu}\bar{\nu}}^{a}(%
\bar{\psi}T^{a}\sigma _{\bar{\mu}\bar{\nu}}\psi )  \nonumber \\
&&+\frac{g}{\Lambda _{L}^{2}}\left[ \tau _{s}\bar{F}_{\bar{\mu}\bar{\nu}%
}^{a}((D_{\bar{\mu}}\varphi )^{\dagger }T^{a}D_{\bar{\nu}}\varphi )+\tau
_{s}^{\prime }g|\varphi |^{2}\bar{F}^{2}\right] .  \label{please}
\end{eqnarray}
where $\mathcal{L}_{1/\alpha }$ is given in (\ref{de}). For simplicity, we
have assumed $U(1)$ charge assignments that forbid terms containing $\bar{%
\psi}^{c}\psi $.

We see that the list of new vertices contains also scalar self-interactions
of type $\varphi ^{6}$, $\varphi ^{4}$-vertices with spatial derivatives,
Pauli terms and several other types of vertices that are not renormalizable
in the framework of the usual power counting.

The couplings $\lambda _{6}$, $\lambda _{4}^{(i)}$, $Y$, $Y^{\prime }$, $%
\tau _{f}$, $\tau _{s}$ and $\tau _{s}^{\prime }$ are weightless. Since $%
\kappa _{i}>0$ their beta functions vanish identically. Following the
arguments explained around formula (\ref{mixec}), the couplings that are not
generated back by renormalization can be consistently switched off, which
produces simplified renormalizable models. The simplest one reads 
\begin{eqnarray*}
&&\frac{1}{2}F_{\hat{\mu}\bar{\nu}}\eta (\bar{\Upsilon})F_{\hat{\mu}\bar{\nu}%
}+\frac{1}{4}F_{\bar{\mu}\bar{\nu}}\tau (\bar{\Upsilon})F_{\bar{\mu}\bar{\nu}%
}+\bar{\psi}\left( \hat{D}\!\!\!\!\slash+\frac{\eta _{f}}{\Lambda _{L}}{\bar{%
D}\!\!\!\!\slash}^{\,2}+\eta _{f}^{\prime }{\bar{D}\!\!\!\!\slash}%
+m_{f}\right) \psi + \\
&&+|\hat{D}\varphi |^{2}+\frac{\eta _{s}}{\Lambda _{L}^{2}}|\bar{D}%
^{2}\varphi |^{2}+\eta _{s}^{\prime }|\bar{D}\varphi |^{2}+m_{s}^{2}|\varphi
|^{2}+\frac{\lambda _{4}\bar{g}^{2}}{4}|\varphi |^{4},
\end{eqnarray*}
which can be cast in $1/\alpha $ form. Again, because $n$ is even the theory
may contain spurious subdivergences.

\paragraph{Theories with four fermion vertices}

Four fermion interactions $\bar{g}_{2}^{2}\bar{\psi}^{2}\psi ^{2}$ are
renormalizable if and only if 
\[
\kappa _{2}\leq 1-\frac{\text{\dj }}{2}, 
\]
which can happen only for \dj $\leq 2$, therefore $4\leq d\leq n+1$. In four
dimensions the simplest case is $n=3$, \dj $=2$, $\kappa _{2}=0$. We can
still choose $\bar{g}_{1}=g$, and, in the presence of scalar fields, $\kappa
_{3}=1/2$. Then the model satisfies (\ref{subd}) and (\ref{b7}), so it is
free of spurious subdivergences.

Turning scalar fields off and choosing $G=SU(N)$, with a Dirac fermion in
the fundamental representation, the lagrangian is the sum of the pure gauge
terms $\mathcal{L}_{Q}+\mathcal{L}_{I}$ plus the fermion kinetic terms, some
Pauli-type terms and the four fermion vertices. Precisely, 
\begin{eqnarray}
\mathcal{L} &=&\mathcal{L}_{Q}+\mathcal{L}_{I}+\bar{\psi}\left( \hat{D}%
\!\!\!\!\slash+\sum_{i=0}^{2}\frac{\eta _{if}}{\Lambda _{L}^{2-i}}{\bar{D}%
\!\!\!\!\slash}^{\,3-i}+m_{f}\right) \psi +\frac{g}{\Lambda _{L}^{2}}{\bar{F}%
}_{\bar{\mu}\bar{\nu}}^{a}\left[ \tau _{f}(\bar{\psi}T^{a}\gamma _{\bar{\mu}}%
\overleftrightarrow{D}_{\bar{\nu}}\psi )+\tau _{f}^{\prime }i\bar{D}_{\bar{%
\nu}}(\bar{\psi}T^{a}\gamma _{\bar{\mu}}\psi )\right]  \nonumber \\
&&+\ \frac{\tau _{f}^{\prime \prime }g}{\Lambda _{L}}i{\bar{F}}_{\bar{\mu}%
\bar{\nu}}^{a}(\bar{\psi}T^{a}\sigma _{\bar{\mu}\bar{\nu}}\psi )+\frac{1}{%
\Lambda _{L}^{2}}\left[ \lambda _{1}(\bar{\psi}\psi )^{2}+\lambda _{2}(\bar{%
\psi}\gamma _{5}\psi )^{2}+\lambda _{3}(\bar{\psi}\gamma _{\hat{\mu}}\psi
)^{2}+\lambda _{4}(\bar{\psi}\gamma _{\bar{\mu}}\psi )^{2}\right]  \nonumber
\\
&&+\ \frac{1}{\Lambda _{L}^{2}}\left[ \lambda _{5}(\bar{\psi}\gamma _{\hat{%
\mu}}\gamma _{5}\psi )^{2}+\lambda _{6}(\bar{\psi}\gamma _{\bar{\mu}}\gamma
_{5}\psi )^{2}+\lambda _{7}(\bar{\psi}\sigma _{\hat{\mu}\bar{\nu}}\psi
)^{2}+\lambda _{8}(\bar{\psi}\sigma _{\bar{\mu}\bar{\nu}}\psi )^{2}\right] .
\label{tre4}
\end{eqnarray}

We have 
\begin{eqnarray*}
\mathcal{L}_{I} &=&\frac{g\lambda _{3}}{\Lambda _{L}^{2}}f_{abc}\tilde{F}_{%
\hat{\mu}\bar{\nu}}^{a}\tilde{F}_{\hat{\mu}\bar{\rho}}^{b}\bar{F}_{\bar{\nu}%
\bar{\rho}}^{c}+\frac{g\lambda _{3}^{\prime }}{\Lambda _{L}^{4}}\bar{D}^{2}%
\tilde{F}^{2}\bar{F}+\frac{g\lambda _{3}^{\prime \prime }}{\Lambda _{L}^{4}}%
\hat{D}\bar{D}\tilde{F}\bar{F}^{2}+\frac{g\lambda _{3}^{\prime \prime \prime
}}{\Lambda _{L}^{4}}\hat{D}^{2}\bar{F}^{3}+\frac{g^{2}\lambda _{4}}{\Lambda
_{L}^{4}}\tilde{F}^{2}\bar{F}^{2} \\
&&+\frac{g}{\Lambda _{L}^{2}}\sum_{m=0}^{3}\lambda _{m}^{(1)}\frac{\bar{D}%
^{2m}}{\Lambda _{L}^{2m}}\bar{F}^{3}{}+\frac{g^{2}}{\Lambda _{L}^{4}}%
\sum_{m=0}^{2}\lambda _{m}^{(2)}\frac{\bar{D}^{2m}}{\Lambda _{L}^{2m}}\bar{F}%
^{4}{}+\frac{g^{3}}{\Lambda _{L}^{6}}\sum_{m=0}^{1}\lambda _{m}^{(3)}\frac{%
\bar{D}^{2m}}{\Lambda _{L}^{2m}}\bar{F}^{5}{}+\frac{g^{4}}{\Lambda _{L}^{8}}%
\lambda ^{(4)}\bar{F}^{6}{}.
\end{eqnarray*}
It is straightforward to check that the $\hat{\partial}$-structures of the
theories listed so far agrees with the results of section 3.

The model (\ref{tre4}) is fully consistent. In particular, it is free of
spurious subdivergences. It is straightforward to include scalar fields and
two scalar-two fermion interactions.

\paragraph{Abelian strictly renormalizable theories}

We conclude with the analysis of a peculiar class of strictly renormalizable
theories. The quadratic part $\mathcal{L}_{Q}$ of the lagrangian must have 
\[
\eta (\bar{\Upsilon})=\eta _{0}\bar{\Upsilon}^{n-1},\qquad \tau (\bar{%
\Upsilon})=\tau _{0}\bar{\Upsilon}^{2(n-1)},\qquad \xi (\bar{\Upsilon})=\xi
_{0}\bar{\Upsilon}^{n-2}. 
\]
For convenience we can choose a strictly-renormalizable gauge fixing, with $%
\zeta (\bar{\upsilon})=\zeta _{0}\bar{\upsilon}^{n-1}$. The IR analysis of
Feynman diagrams is still dominated by the weighted power counting, however $%
\eta (0)=\tau (0)=0$, so the gauge-field propagator contains additional
denominators $\sim 1/\bar{k}^{2(n-1)}$ in the $\left\langle \bar{A}\bar{A}%
\right\rangle $-sector. The loop integrals over $k$ and the loop
sub-integrals over $\bar{k}$ are IR divergent unless 
\begin{equation}
\text{\dj }>4-\frac{2}{n},\qquad \bar{d}>2(n-1),  \label{stric}
\end{equation}
respectively. The latter condition and $n\geq 2$ imply $\bar{d}\geq 3$. We
have also to require (\ref{h}) and (\ref{h2}), and check the absence of
spurious subdivergences.

If \dj $=4$ the gauge coupling itself is strictly-renormalizable and the
theory can be cast in a $1/\alpha $ form. This case, considered in paper I,
is not guaranteed to be free of spurious subdivergences. On the other hand,
if \dj $<4$ the theory can be strictly-renormalizable only if it is Abelian
and contains vertices constructed with the field strength and its
derivatives. In four dimensions no strictly renormalizable theory with $\hat{%
d}=1$ exists, since \dj\ is smaller then $4$, and (\ref{subd}) and (\ref
{d=42}) imply $n<5/3$. Thus, we have to consider higher dimensional
theories. The conditions (\ref{subd}) and (\ref{stric}) give $d\geq 3n$, but
\dj $<4$ gives also $d\leq 3n$, so we must have $d=3n\geq 6$. However, it is
easy to check that the six-dimensional theory with $n=2$, $\bar{d}=5$, \dj $%
=7/2$, is trivial, since no strictly renormalizable interaction can be
constructed. Then we have the nine dimensional theory with $n=3$, $\bar{d}=8$%
, \dj $=11/3$, and lagrangian 
\[
\mathcal{L}=\mathcal{L}_{Q}+\frac{\lambda }{\Lambda _{L}^{20}}\bar{D}^{2}%
\bar{F}^{6}. 
\]
However, since this theory is odd-dimensional, at present we cannot
guarantee that it is free of spurious subdivergences.

\section{More general Lorentz violations}

\setcounter{equation}{0}

So far we have broken the Lorentz group $O(1,d-1)$ into the product of two
subfactors $O(1,\hat{d}-1)\times O(\bar{d})$, which means, for $\hat{d}=1$,
that we have preserved time reversal, parity and rotational invariance. It
is of course possible to break also such symmetries, but that breaking is
not going to affect the results of our present investigation. The structure
of the theory with respect to the weighted power counting is unmodified as
long as each space coordinate has the same weight.

A more general possibility is to break the Lorentz group into the product of
more subfactors, so that different space coordinates may have different
weights. Invariance under spatial rotations is necessarily lost. To cover
the most general case, we can break the spacetime manifold $M$ into a
submanifold $\hat{M}$ of dimension $\widehat{d}$, containing time, and $\ell 
$ space submanifolds $\bar{M}_{i}$ of dimensions $\overline{d}_{i}$, $%
i=1,\ldots \ell $: 
\begin{equation}
M=\hat{M}\times \prod_{i=1}^{\ell }\bar{M}_{i}.  \label{sub}
\end{equation}
Denote the space derivatives of the $i$th space subsector by $\overline{%
\partial }_{i}$ and assume that they have weights $1/n_{i}$. We can assume
also $n_{1}<n_{2}\cdots <n_{\ell }$. Then the weighted dimension \dj , which
is also the weight of the momentum-space integration measure d$^{d}p$, is
equal to 
\[
\text{\dj }=\widehat{d}+\sum_{i=1}^{\ell }\frac{\overline{d}_{i}}{n_{i}}. 
\]
Again, \dj\ can be at most 4. If the super-renormalizable subsector is
non-trivial $d$ must at least be equal to $4$, otherwise Feynman diagrams
can have IR divergences.

The weight of a scalar field $\varphi $ is still \dj $/2-1$, because its
kinetic term must contain $(\hat{\partial}\varphi )^{2}$. Similarly, the
weight of a fermion is \dj $/2-1/2$ and the weight of the $\hat{A}$%
-component of the gauge field is \dj $/2-1$. Since $\bar{\partial}_{i}\hat{A}
$ and $\hat{\partial}\bar{A}_{i}$ belong to the same field-strength
component, the weight of $\bar{A}_{i}$ is \dj $/2-2+1/n_{i}$.

Every argument of this paper can be generalized straightforwardly to the
breaking (\ref{sub}), except for the analysis of spurious subdivergences,
which is a more delicate issue. The conditions $\hat{d}=1$, $d=$even,
combined with suitable other restrictions, are still sufficient to ensure
that no spurious subdivergences occur in the $\hat{k}$-subintegrals. Now,
however, the propagator behaves irregularly also when $\bar{k}%
_{i}\rightarrow \infty $ for any $i<\ell $.

Consider for example a three-factor splitting. The quadratic part of the
lagrangian is a quadratic form in the momenta $\hat{k}$, $\bar{k}_{1}$ and $%
\bar{k}_{2}$, and contains appropriate polynomial functions of $\bar{k}%
_{1}^{2}$, or $\bar{k}_{2}^{2}$, or both. In particular, the $\bar{A}_{2}$%
-quadratic term has the form, in momentum space, 
\[
\bar{A}_{2\mu }(-k)Q_{22}(k)\bar{A}_{2\mu }(k)+\bar{A}_{2\mu }(-k)\bar{k}%
_{2\mu }Q_{22}^{\prime }(k)\bar{k}_{2\nu }\bar{A}_{2\nu }(k), 
\]
where $Q_{22}(k)$ is a polynomial of weight $4-2/n_{2}$. The propagator $%
\langle \bar{A}_{2}\bar{A}_{2}\rangle $ reads 
\[
\langle \bar{A}_{2}(k)\bar{A}_{2}(-k)\rangle =Q_{22}^{-1}(k)\bar{\delta}%
_{2}+P_{22}(k)\bar{k}_{2}\bar{k}_{2}, 
\]
for some unspecified function $P_{22}(k)$. The weight of $Q_{22}$ cannot be
saturated just by $\bar{k}_{1}$, because $Q_{22}\sim (\bar{k}_{1}^{2})^{X}$
would give 
\[
X=2n_{1}-\frac{n_{1}}{n_{2}}, 
\]
which is not integer. Consequently the propagator $\langle \bar{A}_{2}\bar{A}%
_{2}\rangle $ cannot behave regularly in the limit $\bar{k}_{1}\rightarrow
\infty $ with $\hat{k}$ and $\bar{k}_{2}$ fixed, so the $\bar{k}_{1}$%
-subintegrals may contain spurious subdivergences of new types. If we assume 
$\bar{d}_{1}=1$ (in addition to $\hat{d}=1$) then the $\bar{k}_{1}$%
-integrals do not have spurious divergences, as explained in section 4.
However, this is not enough, because the $\hat{k}$-$\bar{k}_{1}$%
-subintegrals themselves, which are two dimensional, can generate spurious
subdivergences. In this case the arguments of section 4 do not apply. We do
not know at present if the problem of spurious subdivergences can be solved
in general. Our present results seem to suggest that the unique consistent
spacetime splitting in the one into space and time.

Before concluding this section it is worth to emphasize that the models to
which our proofs of renormalizability, or absence of spurious
subdivergences, do not apply cannot be completely excluded. Some of them
might work because of unexpected cancellations, which can occur because of
symmetries (e.g. supersymmetry) or peculiar types of expansions or
resummations (e.g. large $N$).

\section{Conclusions}

\setcounter{equation}{0}

In this paper we have completed the program of constructing and classifying
the Lorentz violating gauge theories that are renormalizable by weighted
power counting. The theories contain higher space derivatives, but no higher
time derivatives. We have shown that it is possible to renormalize vertices
that are non-renormalizable in the usual power counting framework, such as
the two scalar-two fermion interactions and the four fermion interactions.
We have studied the time-derivative structure of the theories and the
absence of spurious subdivergences in detail. Spacetime is split into space
and time.

We recall that once Lorentz symmetry is violated at high energies, its low
energy recovery is not automatic, because renormalization makes the
low-energy parameters run independently. One possibility is that the Lorentz
invariant surface is RG stable (see \cite{nielsen} and \cite{colladay}).
Otherwise, a suitable fine-tuning must be advocated.

\vskip 20truept \noindent {\Large \textbf{Acknowledgments}}

\vskip 10truept

I am grateful to P. Menotti and M. Mintchev for useful discussions. I thank
the referee for stimulating remarks.

\vskip 20truept \noindent {\Large \textbf{Appendix A: Propagators and
dispersion relations}}

\vskip 10truept

\renewcommand{\theequation}{A.\arabic{equation}} \setcounter{equation}{0}

After integrating $B^{a}$ out, the gauge-field quadratic terms are contained
in 
\begin{equation}
\mathcal{L}_{Q}+\frac{1}{2\lambda }(\mathcal{G}^{a})^{2},  \label{gu}
\end{equation}
which gives the propagator 
\begin{equation}
\langle A(k)\ A(-k)\rangle =\left( 
\begin{array}{cc}
\langle \hat{A}\hat{A}\rangle & \langle \hat{A}\bar{A}\rangle \\ 
\langle \bar{A}\hat{A}\rangle & \langle \bar{A}\bar{A}\rangle
\end{array}
\right) =\left( 
\begin{array}{cc}
u\hat{\delta}+s\hat{k}\hat{k} & r\hat{k}\bar{k} \\ 
r\bar{k}\hat{k} & v\bar{\delta}+t\bar{k}\bar{k}
\end{array}
\right) ,  \label{pros}
\end{equation}
with 
\begin{eqnarray*}
u &=&\frac{1}{D(1,\eta )},\qquad s=\frac{\lambda }{D^{2}(1,\zeta )}+\frac{-%
\hat{k}^{2}+\zeta \left( \frac{\zeta }{\eta }-2\right) \bar{k}^{2}}{D(1,\eta
)D^{2}(1,\zeta )},\qquad r=\frac{\lambda -\frac{\zeta }{\eta }}{%
D^{2}(1,\zeta )}, \\
v &=&\frac{1}{D(\tilde{\eta},\tau )},\qquad t=\frac{\lambda }{D^{2}(1,\zeta )%
}+\frac{\left( \frac{\tilde{\tau}}{\eta }-2\zeta \right) \hat{k}^{2}-\zeta
^{2}\bar{k}^{2}}{D(\tilde{\eta},\tau )D^{2}(1,\zeta )},
\end{eqnarray*}
where 
\[
D(x,y)\equiv x\hat{k}^{2}+y\bar{k}^{2},\qquad \tilde{\eta}=\eta +\frac{\bar{k%
}^{2}}{\Lambda _{L}^{2}}\xi ,\qquad \tilde{\tau}=\tau +\frac{\hat{k}^{2}}{%
\Lambda _{L}^{2}}\xi , 
\]
and now $\eta $, $\tau $, $\xi $ and $\zeta $, as well as $x$ and $y$, are
functions of $\bar{k}^{2}/\Lambda _{L}^{2}$. The ghost propagator is 
\begin{equation}
\frac{1}{D(1,\zeta )}.  \label{prosg}
\end{equation}

In the Feynman gauge $\lambda =1$, $\zeta =\eta $ we have 
\begin{equation}
u=\frac{1}{D(1,\eta )},\qquad s=r=0,\qquad v=\frac{1}{D(\tilde{\eta},\tau )}%
,\qquad t=\frac{\tilde{\tau}-\eta ^{2}}{\eta D(\tilde{\eta},\tau )D(1,\eta )}%
.  \label{fg2}
\end{equation}

The physical degrees of freedom can be read in the Coulomb gauge $\bar{%
\partial}\cdot \bar{A}^{a}=0$, which can be reached taking the limit $\zeta
\rightarrow \infty $ in (\ref{pros}) and (\ref{prosg}).$\ $In such a gauge
the ghosts are non-propagating, the $\hat{A}$-sector propagates $\hat{d}-1$
physical degrees of freedom with energies 
\[
E=\sqrt{\mathbf{\hat{k}}^{2}+\bar{k}^{2}\eta (\bar{k}^{2}/\Lambda _{L}^{2})} 
\]
and the $\bar{A}$-sector propagates $\bar{d}-1$ physical degrees of freedom
with energies 
\[
E=\sqrt{\mathbf{\hat{k}}^{2}+\bar{k}^{2}\frac{\tau (\bar{k}^{2}/\Lambda
_{L}^{2})}{\tilde{\eta}(\bar{k}^{2}/\Lambda _{L}^{2})}}. 
\]

\vskip 20truept \noindent {\Large \textbf{Appendix B: Renormalizability to
all orders}}

\vskip 10truept

\renewcommand{\theequation}{B.\arabic{equation}} \setcounter{equation}{0}

In this appendix we study the renormalizability of Lorentz violating gauge
theories to all orders, using the Batalin-Vilkovisky formalism \cite{batalin}%
. For simplicity we concentrate on pure gauge theories and use the minimal
subtraction scheme and the dimensional-regularization technique. Note that
the functional integration measure is automatically BRST\ invariant.

The fields are collectively denoted by $\Phi ^{i}=(A_{\mu }^{a},\overline{C}%
^{a},C^{a},B^{a})$. Add BRST sources $K_{i}=(K_{a}^{\mu },K_{\overline{C}%
}^{a},K_{C}^{a},K_{B}^{a})$ for every field $\Phi ^{i}$ and extend the
action (\ref{basis}) as 
\begin{equation}
\Sigma (\Phi ,K)=\mathcal{S}(\Phi )-\int \mathrm{d}^{d}x\left[ \left(
sA_{\mu }^{a}\right) K_{a}^{\mu }+\left( s\overline{C}^{a}\right) K_{%
\overline{C}}^{a}+\left( sC^{a}\right) K_{C}^{a}+\left( sB^{a}\right)
K_{B}^{a}\right] ,  \label{acca}
\end{equation}
Define the antiparenthesis 
\begin{equation}
(X,Y)=\int \mathrm{d}^{d}x\left\{ \frac{\delta _{r}X}{\delta \Phi ^{i}(x)}%
\frac{\delta _{l}Y}{\delta K_{i}(x)}-\frac{\delta _{r}X}{\delta K_{i}(x)}%
\frac{\delta _{l}Y}{\delta \Phi ^{i}(x)}\right\} .  \label{antipar}
\end{equation}
BRST\ invariance is generalized to the identity 
\begin{equation}
(\Sigma ,\Sigma )=0,  \label{nil}
\end{equation}
which is a straightforward consequence of (\ref{acca}), the gauge invariance
of $\mathcal{S}_{0}$ and the nilpotency of $s$. Define also the generalized
BRST operator 
\begin{equation}
\sigma X\equiv (\Sigma ,X),  \label{sigma}
\end{equation}
which is nilpotent ($\sigma ^{2}=0$), because of the identity (\ref{nil}).

The generating functionals $Z$, $W$ and $\Gamma $ are defined, in the
Euclidean framework, as 
\begin{eqnarray}
Z[J,K] &=&\int \mathcal{D}\Phi \exp \left( -\Sigma (\Phi ,K)+\int \Phi
^{i}J_{i}\right) =\text{e}^{W[J,K]},  \label{zj} \\
\Gamma [\Phi _{\Gamma },K] &=&-W[J,K]+\int \Phi _{\Gamma }^{i}J_{i},\qquad 
\text{where\qquad }\Phi _{\Gamma }^{i}=\frac{\delta _{r}W[J,K]}{\delta J_{i}}%
.  \nonumber
\end{eqnarray}
Below we often suppress the subscript $\Gamma $ in $\Phi _{\Gamma }$.
Performing a change of variables 
\begin{equation}
\Phi ^{\prime }=\Phi +\theta s\Phi ,  \label{chv}
\end{equation}
in the functional integral (\ref{zj}), $\theta $ being a constant
anticommuting parameter, and using the identity (\ref{nil}), we find 
\begin{equation}
(\Gamma ,\Gamma )=0.  \label{milpo}
\end{equation}

A canonical transformation of fields and sources is defined as a
transformation that preserves the antiparenthesis. It is generated by a
functional $\mathcal{F}(\Phi ,K^{\prime })$ and reads 
\[
\Phi ^{i\ \prime }=\frac{\delta \mathcal{F}}{\delta K_{i}^{\prime }},\qquad
K_{i}=\frac{\delta \mathcal{F}}{\delta \Phi ^{i}}. 
\]

As usual, renormalizability is proved inductively. The inductive assumption
is that up to the $n$-th loop included the divergences can be removed
redefining the physical parameters $\alpha _{i}$ contained in $\mathcal{S}%
_{0}$ and performing a canonical transformation of the fields and the BRST
sources. Call $\Sigma _{n}$ and $\Gamma ^{(n)}$ the action and generating
functional renormalized up to the $n$-th loop included. The inductive
assumption ensures that $\Sigma _{n}$ and $\Gamma ^{(n)}$ satisfy (\ref{nil}%
) and (\ref{milpo}), respectively.

Locality and (\ref{milpo}) imply that the ($n+1$)-loop divergences $\Gamma
_{n+1\ \text{div}}^{\ (n)}$ of $\Gamma ^{(n)}$ are local and $\sigma $%
-closed, namely $\sigma \Gamma _{n+1\ \text{div}}^{\ (n)}=0$. We have to
find the most general solution to this cohomological condition. In Lorentz
invariant theories the problem has been solved for local functionals with
arbitrary ghost number \cite{barnich}. Since Lorentz invariance does not
play a major role in those proofs, we conjecture that the Lorentz invariant
result generalizes naturally to Lorentz violating theories, namely that $%
\Gamma _{n+1\ \text{div}}^{\ (n)}$ can be decomposed as 
\begin{equation}
\Gamma _{n+1\ \text{div}}^{\ (n)}=\int \mathrm{d}^{d}x\left( \mathcal{G}%
_{n}(A)+\sigma \mathcal{R}_{n}\right) ,  \label{counter}
\end{equation}
where $\mathcal{G}_{n}(A)$ is gauge-invariant.

The functional $\mathcal{G}_{n}$ is local, gauge-invariant, constructed with 
$A$ and its derivatives, and has weight \dj . Since, by assumption, $%
\mathcal{S}_{0}$ contains the full set of such terms, $\mathcal{G}_{n}$ can
be reabsorbed renormalizing the physical couplings $\alpha _{i}$ contained
in $\mathcal{S}_{0}$. We denote these renormalization constants by $%
Z_{\alpha _{i}}$. On the other hand, the counterterms $\sigma \mathcal{R}%
_{n} $ are reabsorbed by the canonical transformation 
\begin{equation}
\mathcal{F}_{n}(\Phi ,K^{\prime })=\int \mathrm{d}^{d}x\sum_{i}\Phi
^{i}K_{i}^{\prime }-\mathcal{R}_{n}(\Phi ,K^{\prime }).  \label{iui}
\end{equation}

Concluding, the ($n+1$)-loop divergences are renormalized redefining the
physical couplings $\alpha _{i}$ and performing a canonical transformation
on the fields and the BRST\ sources. Such operations preserve the identities
(\ref{nil}) and (\ref{milpo}), which are therefore promoted to all orders.

\end{document}